\documentclass[11pt,twoside]{article}
\usepackage{asp2004}
\usepackage{epsf}
\usepackage{psfig}
\usepackage{natbib}

\newcommand{\ie}{{\rm i.e.\ }}
\newcommand{\eg}{{\rm e.g.\ }}

\begin{document}

\title{The Origin and Properties of
X-ray-emitting Gas in the Halos of both Starburst and Normal Spiral Galaxies}
\author{David K.~Strickland}
\affil{Department of Physics \& Astronomy, Johns Hopkins University, 3400 N.~Charles St., Baltimore, MD 21218, USA.}

\begin{abstract}
I discuss the empirical properties of diffuse X-ray emitting
gas in the halos of both nearby starburst galaxies and normal 
spiral galaxies, based on high resolution X-ray spectral
imaging with the Chandra X-ray Observatory. Diffuse thermal
X-ray emission can provide us with unique observational
probes of outflow and accretion processes occurring in star-forming
galaxies, and their interaction with the inter-galactic medium.
I consider both
the  spatial distribution of the diffuse X-ray emission in and around
edge-on starburst galaxies with superwinds
(\eg surface brightness profiles, distribution
with respect to H$\alpha$ and radio emission), and its spectral properties
(\eg thermal or non-thermal nature, abundance ratios, temperatures and
soft and hard X-ray luminosities). These results are 
discussed in the context of current
theoretical models of supernova-driven superwinds,
and compared to the more limited data on extra-planar hot gas around
edge-on normal galaxies.

\end{abstract}

\section{Introduction}
\label{sec:dks:intro}

Galaxy formation and evolution, in particular that of star-forming
galaxies, can not understood in terms of isolated galaxies evolving
independently of their environment. Even in the field galaxy population
mass accretion, mass outflow 
and galaxy/galaxy interactions and mergers
play a fundamental role. Merging of stellar systems is important, but
it not the only significant process, and accretion from the 
inter-galactic medium 
(IGM) and gas and metal loss to the IGM can and probably does occur via purely
gaseous processes even outside dense galaxy group and clusters 
\citep[see \eg][]{toft02,tremonti04}.
To observe and quantify these processes 
in action we must observe the disk/halo interfaces, and halos
(the galaxy/IGM interface) of local star-forming galaxies.

The diffuse thermal X-ray emission from gas
with temperatures between $10^{6}$ and $10^{8}$ K is 
a particularly important probe of the conditions within the halos
of star-forming galaxies. 

\begin{itemize}
\item In the \citet{chevclegg} model for galactic superwinds driven by 
multiple core-collapse supernovae (SN) and massive star stellar winds,
the merged, metal-enriched, SN and wind ejecta has a temperature
within the starburst region of $T\sim  10^{8} \epsilon/\beta K$, and thus would
be a source of \emph{faint} hard thermal X-ray emission ($E>2$ keV).
The fraction of SN mechanical energy thermalized is $\epsilon$ ($\la 1$),
while $\beta$ ($\ga 1$)
is the ratio of total mass added to the hot gas 
compared to that from SN and stellar wind ejecta.
\item The warm neutral and ionized gas in superwinds is observed to
have velocities typically in the range $200$ -- 1000 km s$^{-1}$ 
\citep{ham90,heckman2000}.
This material is believed to be embedded within or at the boundary of
a hot, metal-enriched, wind fluid that has even higher velocities 
\citep[\eg][]{ss2000}.
Various processes will thus create soft thermal X-ray emission ($E<2$ keV),
such as strong shocks ($T\sim 3.5\times10^{6} \, 
[v_{\rm shock}/500 {\rm km\,s}^{-1}]^{2}$) or thermal conduction at interfaces
between the warm gas and the hot wind fluid.
\item The depth of Milky-Way-like galaxy gravitational potential wells 
corresponds to a Virial temperature of 
$T_{\rm vir} \sim 2\times10^{6} \, (v_{\rm circ}/230 {\rm km\,s}^{-1})^{2}$ 
K. If gas inflowing from the IGM
passes through an accretion shock, it will be heated to soft X-ray
emitting temperatures. 
\end{itemize}
In non-active galaxies this emission is purely
collisionally excited, and thus diffuse X-ray emission on galactic scales
is only created by mechanical energy return from SNe or stellar winds, or
accretion into deep gravitational potential wells. The complicated 
processes of photo-ionization and
radiative feedback from massive stars, so important at IR, optical and UV 
wavelengths, are negligible at X-ray wavelengths.

The diffuse thermal X-ray emission from star-forming galaxies is a mixture
of continuum (bremsstrahlung and recombination) and line emission processes.
For soft thermal plasmas ($6 \la \log T (K) \la 7$) line emission
dominates the net emission for abundances $Z \ga 0.2 Z_{\odot}$. As 
newly synthesized metals are believed to initially enter the hot phases
of the ISM, X-ray observations will be vital in capturing galactic chemical
evolution and IGM enrichment at work. Unfortunately the
 CCD-spectrometers on Chandra and XMM-Newton have too low
a spectral resolution to resolve the strong line emission
complexes from O, Fe, Ne, Mg and Si from the continuum over the
energy range 0.5 -- 2.0 keV, which complicates absolute element
abundance measurements for such soft metal-bearing plasmas. 
Nevertheless, some useful spectral elemental 
diagnostics are possible with existing
instrumentation, as I shall discuss.

Much of this contribution is based on our recent Chandra-based survey
of 7 edge-on starbursts and 3 edge-on normal spirals 
\citep{strickland04a,strickland04b}, along with a preview of more recent
work on diffuse hard X-ray emission from M82. Our findings on the
starburst systems and their superwinds are presented in 
\S~\ref{sec:dks:starbursts:hard} and \ref{sec:dks:starbursts:soft} 
Their properties are compared
empirically to the diffuse X-ray properties of the normal
spirals in \S~\ref{sec:dks:both}. Conclusions drawn from this
work, and some further questions raised by it, are summarized
in \S~\ref{sec:dks:conclusions}

\section{Normal vs. Starburst}
\label{sec:dks:intro:definitions}

To understand the X-ray properties of normal star forming
galaxies we must use a robust and physically-motivated method for
separating more powerful objects (starbursts) from
normal galaxies with very similar total star formation rates.

We define star-forming galaxies as starbursts using
the widely-used IRAS 60$\micron$ to 100 $\micron$ 
flux ratio $f_{60}/f_{100} \ge 0.4$ \citep[\eg][]{lehnert95}.
There is a continuous distribution in the intensity in disk
and irregular galaxies, but galaxies
above this $f_{60}/f_{100}$ 
thresh-hold have mean star formation intensities considerably
greater than in a normal galaxy such as the Milky Way.
This ratio is a measure of the luminosity-weighted mean dust temperature
in these galaxies, but is well correlated to the inverse of
the gas consumption timescale $\tau_{SF}$ due to star-formation, where
a traditional and physically motivated definition of a starburst is
$\tau_{SF} \le$ a few times $10^{8}$ yr. 

The dust temperature is a function of
the energy density in the interstellar radiation field, 
in particular the
energy density in the UV which is dominated by massive stars in the
presence of moderate to high levels star formation 
\citep[see \eg][]{desert90}. Thus the $f_{60}/f_{100}$
ratio is also a measure of the intensity of SF (\ie rate per unit area
or volume), rather than just the gross rate. It is thus complementary to
other traditional methods of estimating the mean galactic
SF rate per unit area
based on total IR, H$\alpha$ or non-thermal radio luminosities
divided by some effective area \citep[][]{dettmar93,rand96,dahlem01,rossa03a}.

That galaxies found by this method are not just classic \emph{nuclear}
starbursts like the archetypes M82 and NGC 253. 
For example, NGC 4631 ($f_{60}/f_{100}=0.40$, 
see \citealt{rice88}, or $f_{60}/f_{100}=0.53$ from \citealt{sanders03}) 
has only a weak nuclear
starburst \citep{golla94b}, but elevated levels of star formation occur
across a large fraction of the optical-disturbed disk.


The starburst galaxies in our sample are M82, NGC 253, NGC 1482, 
NGC 3079, NGC 3628, NGC 4631 and NGC 4945. Other well-studied
edge-on galaxies fulfilling this starburst definition
are NGC 2146, NGC 3556 and NGC 5775. The three normal
galaxies in our sample are NGC 891 and the lower mass
spirals NGC 4244 and NGC 6503.

Although there are too few normal galaxies in this sample to
robustly determine the diffuse X-ray properties of this
class on its own, we can
determine whether the spectral  and spatial properties of any
diffuse emission in these galaxies is similar to or different from
the larger sample of starburst galaxies with known outflows.
If the hot gas in the halos of normal galaxies is primarily generated
by a mechanism other that SF-related disk outflow then their
soft X-ray properties should be distinct from superwinds.

\section{Hard diffuse X-ray emission in starbursts}
\label{sec:dks:starbursts:hard}

Chandra observations of star forming galaxies have finally
proven that the summed emission from populations of accreting
compact objects (normal X-ray binaries, ultra-luminous X-ray sources, 
and in some cases low-luminosity AGN)
dominate the hard X-ray emission ($E>2$ keV), 
see \eg \citet{swartz03,colbert04}.  The luminosity function 
of these X-ray sources is such that the brightest point sources account
for the majority of the net emission, so that Chandra observations
of nearby galaxies typically resolve out $\sim90$\% of the total
hard X-ray emission. The remainder is consistent with unresolved
point source emission based on extrapolation from the observed luminosity 
function.

\begin{figure}[!t]
  \plotone{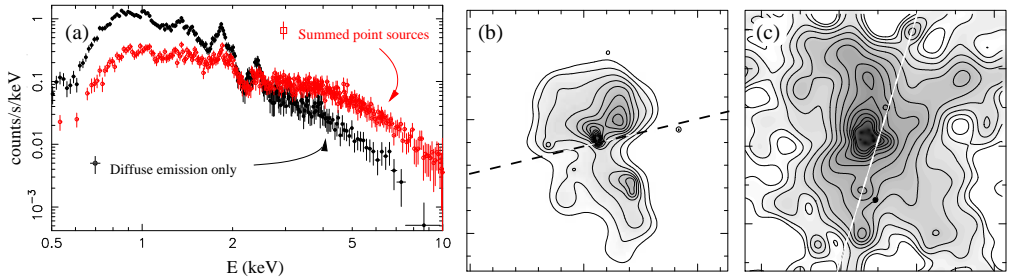}
  \caption[]{(a) Chandra ACIS-S3 spectra of the central 1 kpc of M82,
  separated in to the diffuse emission spectrum with weak Fe K-shell 
  emission, and the summed point source
  spectrum. From Strickland et.~al, in preparation.
  (b) A smoothed image of
  E=0.3-2.0 keV diffuse X-ray emission from NGC 3628, shown on a log
  scale, contours increasing by factors of $\surd 2$. The size of the
  image corresponds to a $20\times20$ kpc region. The dashed line shows the
  mid-plane of the galaxy disk. 
  Note the spurs of soft X-ray emission at the edges
  of the superwind.
  (c) As b, except for NGC 3079, shown on the same surface brightness
  scale. Adapted from \citet{strickland04a}.
  }
  \label{fig:dks:spectrum_image}
\end{figure}

Nevertheless, Chandra observations have shown that in a few 
starburst galaxies there is genuine diffuse hard X-ray emission
within the starburst region, in addition to the dominant point 
source emission, and in excess of
any unresolved point source contribution. These galaxies are
M82 \citep[][see also Fig.~\ref{fig:dks:spectrum_image}a]{griffiths2000}, 
NGC 253 \citep{weaver02}, and NGC 2146 
(H.~Matsumoto,
unpublished), all intense starbursts with SF rates per unit area 
near the upper limit observed for any starburst \citep{meurer97}.
Studies of diffuse X-ray emission from galaxies have always concentrated
on soft X-ray emission, so these hard X-ray detections may open
a new window on an understudied but important phase of the ISM. 
We intend to perform a more systematic search for fainter 
diffuse hard X-ray emission in Chandra observations of less intense
starbursts, but at present these three galaxies are the best examples
of genuine diffuse hard X-ray emission.

The likely causes for appreciable
diffuse hard X-ray emission in a star-forming galaxy can be divided between
the action of low luminosity AGN (X-ray fluorescence and 
scattered AGN light) and processes associated with
SNe (thermal emission from merged SN ejecta, non-thermal emission from
inverse Compton [IC] scattering of IR photons off cosmic rays). 
In NGC 253 photoionization by the LLAGN is the
most probable cause \citep{weaver02}. M82 has no LLAGN,
and the close spatial association of the diffuse hard X-ray emission
with the starburst region implicates SN activity 
(see Fig.~1 in \citealt{strickland_brazil}).

Is the hard emission in M82 consistent with theoretical models of hot gas in
starburst galaxies? The initial Chandra ACIS-I study of this emission by 
\citet{griffiths2000} emphasized a thermal interpretation of the 
hard continuum and a possible E=6.7 keV line emission from Fe XXV.
As only gas with $\log T \ga 7.5$ would produce appreciable 6.7 keV
Fe emission, this implies $\epsilon/\beta \ga 0.3$, i.e. efficient
SN thermalization. 
However, the observed hard X-ray luminosity ($L_{\rm 2-10 keV}
\sim 5\times10^{39}$ erg s$^{-1}$) is at least an order of magnitude
greater than expected for purely thermal emission from the starburst
region (Strickland et.~al, in preparation).
Our analysis of new ACIS-S data and recalibrated ACIS-I data
finds that the continuum is better fit by a non-thermal power law
model, as expected from IC emission.
Line emission from Fe is weak, but present in both datasets, although spread
over the energy range 6.3 -- 6.8 keV, suggesting emission from both
neutral and highly ionized Fe. The hard X-ray
spectrum is reminiscent of that of the Galactic Ridge \citep[\eg][]{park04}.
Unlike the Galactic Ridge, 
whose energy source is a matter of debate, the
flux in the ionized component of the Fe emission is consistent with
predictions based on the \citet{chevclegg} model and M82's inferred
SN rate. Although the nature of the diffuse hard X-ray emission 
in the center of
M82 is more complicated than initially thought, highly
ionized ($\log T \ga 7.5$) metal-bearing gas exists at the 
luminosity expected from superwind models, in the presence
of a brighter but non unexpected non-thermal diffuse continuum source.
With the line-based spectral diagnostics possible with the
Astro-E2 calorimeter ($E/\Delta E \sim 1000$ at 6 keV) 
it should be possible to measure the
temperature of the 6.7 keV-emitting material independently of 
the non-thermal continuum, and hence directly constrain the efficiency
of supernova feedback in a starburst.

\section{Extra-planar soft diffuse X-ray emission from superwinds}
\label{sec:dks:starbursts:soft}

There is a long history of soft X-ray studies of starburst
galaxies with superwinds, leading up to modern
$1\arcsec$ spatial resolution studies with Chandra.

The diffuse soft X-ray luminosity is very well
correlated with the star-formation rate 
($\log L_{\rm X,TOT}/L_{\rm FIR} \approx -3.6\pm{0.2}$
and $\log L_{\rm X,HALO}/L_{\rm FIR} 
\approx -4.4\pm{0.2}$ [e.g. Fig.~\ref{fig:dks:plots}a], 
where the halo is defined as the region $|z|>$2 kpc
from the mid-plane), in good  agreement with theoretical
expectations \citep{strickland_brazil}.

Typically the soft diffuse X-ray emission is most extended 
along the minor axis of the host galaxy, and is spatially correlated
with H$\alpha$ emission. The maximum height to which emission
is detected depends on the size of the host galaxy 
\citep[Grimes et.~al, in preparation]{strickland04b}. In dwarf starbursts
$z_{max} \sim 2$ kpc 
\citep[\eg][]{martin02,ott02_thesis}. In more typical
local starbursts ($\log L_{FIR} [L_{\odot}] 
\sim 10.5$) $z_{\rm max} \sim 10$ -- 20 kpc,
while $z_{max} \sim 10$ -- 50 kpc in ULIRGs.
In the edge-on $\log L_{FIR}\sim10.5$ starbursts the minor-axis 
X-ray surface brightness profiles are best fit by exponential
models with scale heights  $H_{\rm eff}\sim$ 
2 -- 4 kpc (density scale height 4 -- 8 kpc). The data is inconsistent
with the $\Sigma \propto z^{-3}$ power law expected of a
simple radial or conical volume-filling flow.

The emission is thermal, with clear metal line emission features,
and typically has a characteristic temperature in the 
range 2 -- 8 million degrees. Care needs to be taken when fitting to
spectra containing spectrally distinct regions \citep{weaver00}.
Even apparently spectrally uniform regions are not well
described as single temperature plasmas in ionization
equilibrium \citep{strickland04a}.  Conservative analyzes of
the spectra do not allow absolute element abundances to be determined,
but relative abundances, such as O or other $\alpha$-elements with
respect to Fe, are well constrained \citep{martin02,strickland04a}. 
The super-Solar
$\alpha$/Fe ratios observed are consistent with
both massive star enrichment, or residual Fe depletion on dust (dust
destruction time scales in the hot gas are comparable to wind flow times).

The spectral properties of the diffuse emission do vary
spatially along the wind. Primarily this is due to variations in
intervening absorption column (by a factor $\ga 10$)
in these edge-on galaxies, when moving
from emission within the absorbed disk plane 
to extra-planar emission. There
appears to be a weaker (factor $\sim 2$) drop in effective temperature
from disk to halo. Spectral variation within the halo is generally
negligible, although weak variations with $z$ are see in a few cases
\citep{strickland04a}. The combination of spectral and surface
brightness variations are not consistent with 
the adiabatic expansion of a volume-filling X-ray emitter
\citep[\eg][]{sps97}.

In all cases with good signal to noise data, the
Chandra observations reveal
genuine spatial structure in the superwind 
soft X-ray emission (Fig.~\ref{fig:dks:spectrum_image}), on scales similar
to the structure in optical nebular emission 
\citep{strickland00,strickland02,cecil02,schurch02,strickland04a}.
The best interpretation of this data is in terms of 
the soft X-ray emission arising in low volume regions in the vicinity
of obstacles in the flow: the walls, and any dense clouds within the wind.
The large spurs and filaments are mainly the limb-brightened walls of
the outflow.

The combination of spatial structure, close relationship to optical nebular 
emission, exponential surface brightness profiles, 
weak spectral variation and high apparent gas phase $\alpha$/Fe
abundance ratios force the following conclusions regarding the soft
X-ray emitting material in superwinds:
It is not a volume-filling wind of
the kind modeled by \citet{suchkov96}. The majority of the emission
must come from relatively low volume filling factor, 
$0.01 \la \eta \la 0.3$ (more work needs to be 
done to better quantify these values). It may only contain a small
fraction of the energy and metal content of the wind.
The soft X-ray arise in some form of interaction between dense ambient
gas in clouds or at the outflow walls and the volume-filling
tenuous merged SN ejecta fluid that really drive the superwind.
Much of our current work is exploring  methods
for distinguishing between the multiple plausible wind/ambient ISM interaction
models (shock-heated ambient gas, conductively evaporated and
heated ambient gas, or shock-compressed ejecta fluid) 
implied by the Chandra data.

\section{Diffuse X-ray emission in both normal and starburst galaxies}
\label{sec:dks:both}

\begin{figure}[!t]
\plotone{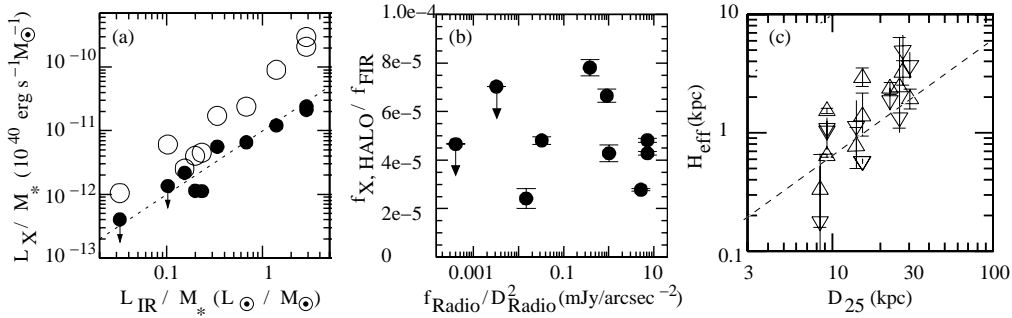}
\caption{(a) Disk and halo diffuse X-ray luminosity per unit galaxy
  mass, as a function of FIR luminosity, effectively the total SF rate
  (disk [$|z|< 2$ kpc]: 
  open circles, halo: filled circles).
  (b) Halo-region diffuse X-ray flux to host galaxy FIR flux ratio,
  as a function of 
  the mean SF rate per unit area in the host galaxy.
  (c) Minor axis surface brightness scale height $H_{\rm eff}$ as
  a function of the host galaxy optical diameter $D_{25}$.
  In both plots a and c the dashed line 
  is not a fit, and is shown purely to show what unit slope is.}
\label{fig:dks:plots}
\end{figure}

We detect extra-planar ($|z| > 2$ kpc) hot gas in only one of 
the three normal galaxies in \citet{strickland04a,strickland04b}: NGC 891,
which  has a similar $L_{\rm X, HALO}/L_{\rm FIR}$ ratio to the 
starbursts (Fig.~\ref{fig:dks:plots}a). This is interesting, as one
might expect
 more halo X-ray emission from galaxies with more intense star
formation.
However, both NGC 4244 and NGC 6053 have lower star formation rates
than NGC 891, and they could plausibly have hot halos at luminosities
consistent with this relationship that would not have been detected
in these observations.

The minor axis extent 
of the diffuse X-ray
emission (measured by scale height or with an isophotal size) 
scales with the size of the host galaxy (e.g. optical $D_{25}$ or
K-band half light radius). Both NGC 891 and NGC 6503 are consistent
with the starburst trends. The size of the star forming region
within the disk has no influence on the minor axis
extent of the X-ray-emitting gas.

We find reasonably good correlations between the mean and halo-region
X-ray surface brightness and mean SF rate/area in the disk (based
on $f_{60}/f_{100}$, $L_{\rm FIR}/D_{25}^{2}$ 
or non-thermal radio flux and size 
$f_{\rm radio}/D_{\rm radio}^{2}$). This is similar to the
qualitative correlation between the prevalence of extra-planar
H$\alpha$ emission or dust with mean SF rate per unit 
area \citep{dettmar93,rand96,howk99b}.
 It is tempting to interpret our X-ray results
in terms of a critical SF (or SN) rate per unit area required for
superbubble blow out from a disk galaxy. However, this correlation
is possibly a combination of the more fundamental and separate
X-ray flux to 
total SN rate (e.g. $L_{\rm X, HALO}/L_{\rm FIR}$) and X-ray
size to galaxy size correlations. 

\section{Conclusions and remaining questions}
\label{sec:dks:conclusions}

Chandra observations of starburst galaxies
have answered some of the pressing questions regarding
X-ray emitting gas in superwinds. The very hot plasma predicted by the 
\citet{chevclegg} model has probably been detected in M82 at expected
flux levels in the 6.7 keV Fe line,
along with non-thermal continuum emission. Soft X-ray emission
in winds is of low volume filling factor, as is associated with
some form of wind fluid interaction with cooler denser ambient disk
and halo gas. Strongly mass-loaded wind models can be ruled out.

The extra-planar X-ray luminosity of the approximately edge-on
star-forming disk galaxies
in our sample is directly proportional to the 
host galaxy's SF rate. The vertical extent of the diffuse X-ray emission
is proportional to the host
galaxy's size.
We definitely need more X-ray observations
of edge-on \emph{normal} star-forming disk galaxies, especially if we are
to constrain accretion models. What we can say
with the existing observations is that the properties of the diffuse
soft X-ray-emitting plasma in normal galaxies appears very similar to
that in starbursts, scaled proportionally to lower SF rates.
The existing non-detections of extra-planar hot gas do not prove 
that there is no
hot halo gas.

With the aim of being provocative, I will close by asking some questions
inspired by these results. How similar can the physical conditions in 
a normal galactic fountain be to a superwind? Can warm and hot gas
filling factors be low in the halo of NGC 891, despite the apparent
smoothness of the emission? Are the majority of non-detections of
extra-planar warm ionized gas, hot ionized gas, radio emission, and
dust significant? In other words, is there no extra-planar plasma, 
or is it just too faint to detect?

\acknowledgements

It is a pleasure to thank the organizers for such a 
stimulating and well-run meeting. I owe thanks to my collaborators,
in particular Ed Colbert, 
Michael Dahlem, John Grimes,  Jo Hartwell, Tim Heckman, Charles Hoopes,
Trevor Ponman, Andy Ptak, Ian Stevens, Leslie
Summers, and Kim Weaver.  This work was funded by NASA through SAO grant
PF010012.


\end{document}